\documentstyle[preprint,aps]{revtex}
\begin{document}
\draft

\title{Theory of Self-organized Criticality for Problems with 
Extremal Dynamics}
\pagestyle{myheadings}
\markboth{Gabrielli, Cafiero Marsili and Pietronero}{Gabrielli, 
Cafiero Marsili and Pietronero}
\author{A.Gabrielli$^{1,2}$, R. Cafiero$^2$, M.Marsili$^3$ and  
L.Pietronero$^2$ }
\address{
$^1$Dipartimento di Fisica, Universit\'a di Roma "Tor Vergata", Via 
della
Ricerca Scientifica 1, I-00133 Roma}
\address{
$^2$Dipartimento di Fisica, Universit\'a di Roma "La Sapienza", 
P.le Aldo Moro 2, I-00185 Roma, Italy; and INFM, unit\'a di Roma I}
\address{
$^3$Institut de Physique Theorique Universit\`e de Fribourg, 
P\'erolles 
CH-1700 Fribourg, Suisse}
\maketitle
\date{\today}
\maketitle
\begin{abstract}
We introduce a general theoretical scheme for a class of
phenomena characterized by an extremal dynamics and 
quenched disorder.
 The approach is based on a transformation of the 
 quenched dynamics into a stochastic one with cognitive 
 memory and on other concepts which permit a 
mathematical characterization of the self-organized
 nature of the avalanche type dynamics. In addition it is 
possible to compute the
relevant critical exponents directly from the microscopic 
model. A specific application 
to Invasion Percolation is presented
  but the approach can be easily extended to various other 
problems.
\end{abstract}
\smallskip
{\small PACS numbers: 02.50.-r;05.40.+j;05.90.+m}
\newpage

Extremal models of self-organized criticality (SOC \cite{Bak}) 
have recently attracted significant 
theoretical attention. This class of models
describes phenomena ranging from fluid displacement in porous 
disordered media \cite{Wilk}, to punctuated biological evolution
\cite{Bak}. In these models, at each time 
step the dynamical activity is concentrated
on the site with the {\em extremal} value 
of a quenched disordered variable. This rule leads
 to a rich and complex behaviour, which 
 has been widely studied \cite{gap}.
 
In this letter we describe a general 
theoretical approach which addresses 
the basic problems of extremal models: 
(i) the understanding of the scale-invariance and 
self-organization;(ii) the origin of the avalanche
dynamics and (iii) the computation of 
the relevant critical exponents. We apply it
specifically to Invasion Percolation (IP), but
it can be easily extended to other models 
of this type like the Bak and Sneppen
model \cite{BSRG}.

The usual real space methods, like Fixed Scale Transformation
\cite{FST2}, cannot address directly 
the problem of the irreversible dynamics
with {\em quenched} disorder. 
In order to overcome this
basic problem we introduced a mapping of 
a quenched extremal dynamics into a stochastic 
one with cognitive memory, the {\em 
quenched-stochastic transformation}, also called
 Run Time Statistics (RTS) \cite{RTS}. 
This approach was improved in various
steps \cite{46matt,RTS,BSRG} and now we can 
develop it into a general
theoretical scheme. Its essential points are:
\\
- Quenched-stochastic transformation.
\\
- Identification of the microscopic fixed point
dynamics (SOC).
\\
- Identification of the scale invariant dynamics 
for block variables.
\\
- Definition of {\em local} growth rules for the extremal model. This 
clarifies the origin of {\sl avalanche
dynamics}.
\\
- Use of the above elements in the FST 
scheme to compute analytically the 
relevant exponents of the model.

Let us start with the quenched-stochastic 
transformation. We will discuss it 
for the case of bond Invasion Percolation
 (IP) \cite{Wilk}. The IP model describes the capillary displacement of 
a fluid in a porous medium. The medium is represented as a network 
of bonds. To each bond $i$ is assigned a quenched random number 
$x_i$ extracted from a given distribution with density $\rho(x)=1$. 
The invading cluster evolves by occupying the bond with the smallest 
$x_i$ on its perimeter. The basic idea is to
 map the deterministic extremal IP dynamics 
into an annealed stochastic process. 
 A general stochastic 
process is based on the following elements: a) a set
of time dependent dynamical 
variables $\{ \eta_{i,t} \}$; b) a Growth Probability Distribution 
(GPD) for the single growth step $\{ \mu_{i,t} \}$,
obtained from the $\{ \eta_{i,t} \}$; 
c) a rule for the evolution of the 
dynamical variables $\eta_{i,t}\to\eta_{i,t+1}$. 

Therefore, in order to map IP onto a stochastic process we 
have to identify the above elements.

An insight into the essence of the question is given by
by the following example.
 Consider two independent random variables 
$X_1,X_2$ uniformly distributed in $[0,1]$ and 
let us eliminate the smallest, for example $X_2$. 
The probability
that $X_2<X_1$ is $1/2$. 
At the second ``time step'', we compare the
surviving variable $X_1$ with a third, uniform, random 
variable $X_3$ just added to the game and, again, we 
eliminate the smallest
one. In this case we need to compute the 
probability $\mu_3$ that $X_3<X_1$ {\em given}
that $X_2<X_1$. This, using the rules of conditional 
probability, reads:
$$\mu_3=\tilde{P}(X_3<X_1)=P(X_3<X_1|X_2<X_1)=$$
\begin{equation}
 =\frac{P(X_3<X_1\bigcap X_2<X_1)}{P(X_2<X_1)}=\frac{2}{3},
 \label{eqexamp1}
\end{equation}
where $P(A|B)$ is the probability of
the event $A$, given that $B$ occurred, and 
$P(A\bigcap B)$ is the probability of occurrence of 
both $A$ and $B$. Equation \ref{eqexamp1} tells us that the 
information $X_2<X_1$ 
{\em changes in a conditional way} the {\em effective} 
probability density
$p_1(x)$ of $X_1$. In fact, by imposing the condition 
$X_2< x$ (given $x<X_1<x+dx$) we get $p_1(x)=2x$.
Qualitatively, the event $X_2<X_1$ decreases the
probability that $X_1$ has small values.

The above example contains the essential idea 
of the {\em quenched-stochastic transformation}. 
Let us come back to the IP model. In view of the above example, 
each perimeter variable will have, at time $t$, an effective density 
depending on the past growth history of the bond. If a variable has 
lost many times, it will have a density more and more concentrated on 
great values. The past history of a variable can be represented by only 
one parameter, the number of time steps that the variable has spent 
in the perimeter, that we call "age" $k$ of the variable. Variables with 
the same age $k$ will have the same effective density $p_{k,t}(x)$ 
($p_{0,t}(x)=\rho(x)$) at time $t$. At this point we can express the 
probability that a variable of age $k$ is selected at time $t$, 
i.e. it is the smallest perimeter variable, given the effective 
densities $p_{k,t}(x)$ of all the perimeter variables 
for a realization of the process at time $t$, in the following way:
\begin{equation}
\mu_{k,t}
= \int_{0}^{1}dx\:p_{k,t}(x)\prod_{\theta} 
(1-P_{\theta,t}(x))^{n_{\theta,t}-\delta_{\theta,k}},
\label{mu}
\end{equation}
where $P_{k,t}(x)=\int_0^xdyp_{k,t}(y)$, the 
product is intended over all the ages of the variables
and $n_{\theta,t}$ is the number
 of active variables of {\em age} $\theta$ at time $t$. The 
product inside the integral
takes into account the competition of the
 selected variable with each one of the other active variables, 
while the integral between $0$ and $1$ takes into account 
all the possible values of the growing variable. The temporal 
evolution of the densities of the still active 
variables is then given by:
\begin{equation}
p_{\theta+1,t+1}(x) = p_{\theta,t}(x) \int_{0}^{x}
\frac{m_{k,t}(y)}{1-P_{\theta,t}(y)}dy.
\label{nuova}
\end{equation}
where $m_{k,t}(y)=\left[ p_{k,t}(x) \prod_{\theta}
\left( 1-P_{\theta,t}(x) \right)^{n_{\theta,t}-
\delta_{\theta,k}} \right] / \mu_{k,t}$.

Equations \ref{mu}, \ref{nuova} {\em
 describe a quenched extremal 
process as a stochastic process with memory}. 
The presence of memory is enlighted by the
 dependence of the GPD on $k$. It has been shown by 
M. Marsili \cite{RTS} that a mean field like
expansion of Eq.(\ref{mu}) in the limit $t \to
\infty$ gives: $\mu_{k,\infty} \sim 
\frac{1}{(k+1)^{\alpha}}$. The power
law behaviour of $\mu_{k,t}$ guarantees that
screening is preserved
 at all scales, which is the condition 
 to generate fractal structures \cite{Raf}.

An important quantity to study in IP is the histogram 
$\Phi_t(x)$, which is the distribution at time $t$ of the 
perimeter variables. It has been shown numerically \cite{Wilk} that
 the histogram of IP self-organizes asymptotically 
into a stable shape, a theta function with a discontinuity at
 $x=p_c<1$, where $p_c$ is the critical bond percolation 
threshold. In ref. \cite{RTS} an equation for the temporal evolution of 
$\Phi_t(x)$ is derived analytically directly from the RTS microscopic 
dynamics, without {\em a priori} assumptions. The final form of this 
{\em histogram equation}, obtained after some technical 
approximations \cite{RTS} is:
\begin{equation}
\partial_{x}\Phi_{t}(x)=\beta\Omega_{t}
\Phi_{t}^{2}(x)\:\left[1-\frac{\omega_t}
{\omega_t+1}\Phi_{t}(x)\right]
\label{eqphi}
\end{equation}
where $N_t$ is the number of perimeter variables a time $t$ for a realization of the growth process, $\Omega_t=\langle{N_t}\rangle$, 
$\omega_t=\langle N_{t+1}-N_t \rangle$, the mean $\langle ... \rangle$ 
is over different realizations, and
$\beta$ is the solution of:$\beta=1-e^{-\beta \Omega_t}$.
 The solution of eq. \ref{eqphi} is shown to converge 
 asymptotically to a theta function (fig.\ref{figphi}): 
\begin{equation}
\lim_{t \rightarrow \infty}
\Phi_t(x)=\frac{1}{1-p_c}\theta(x-p_c)
\label{asimp}
\end{equation}
where $p_c=1/2$ for $2-d$ bond IP. This result 
clarifies the SOC nature of the problem and it will 
be crucial in the definition of avalanche dynamics. A 
similar equation describing the SOC 
behaviour of extremal models, 
the {\em gap equation}, has been obtained by Bak et al. 
\cite{gap}, based on phenomenological assumptions. 

The next step is the identification
of the scale invariant dynamics 
for block variables. This point
can be easily addressed. In fig. \ref{fig1} we
 show a coarse graining procedure for the
 extremal dynamics of IP. In the left side
of the figure we show some paths 
 leading to cell $B$ (or $A$), each one composed by a set of 
 quenched variables ($\{\epsilon_i\}_{b}$). Each path is 
characterized by the largest variable in the set 
 ($\epsilon_b$). The best path leading to cell $B$ 
is that with the smallest $\epsilon_b$ (saddle point), and will compete
 with the best path leading to cell $A$. So, we identify the 
 block variables $\epsilon_{A}$ and $\epsilon_B$ 
 with the saddle point of the best path 
 leading to the corresponding cell, say:
\begin{equation} 
\epsilon_B=F_B\left[\{\epsilon_i\}_b\right]=
\min_b\left[\max_i\{\epsilon_i\}_b\right].
\label{coarsegr}
\end{equation}
and analogously for $\epsilon_A$. 
Of course the initial density of the block variables 
$\epsilon_A$ and $\epsilon_B$ will
 be rather complex. However, 
extremal dynamics is not perturbed by a different choice 
of the initial density 
of the variables. In fact, eq. \ref{mu} is invariant under the 
transformation $x \to \int_0^x p_0(y) dy$ which maps the density 
$p_0(x)$ onto the flat one \cite{RTS}. 
So, we conclude that the coarse-grained
 dynamics is {\sl intrinsically scale invariant}.

However, in order to compute 
the critical IP exponents by real space methods 
like FST we need {\em local growth rules}. This corresponds to limit 
the process inside the FST growth column \cite{FST2}, which implies
 a truncation of eqs. \ref{mu}, \ref{nuova} in order to follow the RTS dynamics of only a finite number of variables, instead of those of 
the whole perimeter. This step must be realized carefully, because screening in IP is power law like (we cannot simply ignore what happens outside the growth column), and can be realized in the following way.
Let us consider the asymptotic IP dynamics.
 Both simulations \cite{Roux,Masl} 
 and analytical results \cite{gap}, together with eq. \ref{asimp}
 show that the fixed point IP dynamics 
 develops itself in scale invariant macro events, or 
{\em avalanches}. An avalanche is a {\em spatially and 
causally connected sequence} of single growth events. 
A critical scale invariant avalanche starts with a variable at $p_c$ 
\cite{Roux,gap}. By definition, all other 
 variables in the avalanche 
 have values smaller than $p_c$. In order to 
restrict the stochastic process to a limited
region of the system {\em we can consider 
only the dynamics inside a single avalanche}. In view 
of the above arguments, the RTS equations
  for this {\em local dynamics} are obtained 
  from Eqs. (\ref{mu}, \ref{nuova})
  by taking into account only 
  the variables which become active after the initiator's 
  growth and by integrating in Eq.(\ref{mu})
  only in $[0,p_c]$ and not in $[0,1]$. This will lead to a GPD for the
 avalanche dynamics which is defecting, i.e. with a normalization less 
than one, in that it must account of the probability that the avalanche 
stops \cite{rtslung}.

At this point, we use the scale invariant 
IP local growth rule in the FST scheme, to 
compute the fractal dimension of the infinite 
percolating cluster \cite{rtslung}. 
This is accomplished by computing the transverse 
nearest-neighbour correlations at a generic scale 
via suitable path integrals. 
In Table \ref{table} we show the results of our FST 
calculations for invasion percolation and
for some related models.
The convergence of the path integrals with respect to 
the path length $n$ is power 
law like \cite{RTS}, and we need to extrapolate
the FST results to $n=\infty$. 
The results of our calculations are in very good agreement
 both with numerical simulations and with known analytical
 values (Table \ref{table}).
 
 We can compute also the 
 exponent of the critical avalanche
 distribution of IP. The distribution of off-critical avalanches in 
 Invasion Percolation ($p \neq p_c$) has been shown to have the 
form \cite{gap,Masl}:
\begin{equation}
D(s;p) = s^{-\tau}f(|p-p_c|s^{\sigma})
\label{disval}
\end{equation}
where $s$ is the avalanche size and $p$ is the initiator 
of an avalanche. In the limit $t\rightarrow \infty$ the system 
self-organizes into the critical state $p=p_c$, 
and the (normalized) avalanche size distribution becomes:
\begin{equation}
D(s;p_c)=\frac{s^{-\tau}}{\sum_{s=1}^{\infty}s^{-\tau}}
\label{1valPc}
\end{equation}
Our calculation scheme develops in the following steps:
1) we evaluate the left hand side of Eq. (\ref{1valPc});
2) we solve equation (\ref{1valPc}) for $\tau$. For $s=1$
 we can write:
 \begin{equation}
D(s=1;p_c) = \sum_{j} W^{(j)}\cdot P^{(j)}(s=1;p_c) 
\label{finale}
\end{equation}
where $W^{(j)}$ are the weights of the different boundary 
conditions near the initiator of the avalanche and $P^{(j)}(s=1;p_c)$ 
are the stopping probability of the avalanche after one step, 
computed with the local RTS growth rules.

By Inserting Eq.(\ref{finale}) in Eq.(\ref{1valPc}) we get: 
$\tau\simeq 1.5832$.
Also this analytical result is in excellent agreement with 
numerical simulations \cite{Masl}, which give $\tau=1.60$. 
In order to check independently our 
theory, we have performed 
a new type of computer simulations 
for the avalanche distribution. 
From RTS scheme and the discussion 
 of the scale invariant local dynamics, one notes 
 that $k(t_0+s)$, the age of the bond grown 
during the avalanche at time $t_0+s$, must 
 satisfy the condition $k(t_0+s) < s$.
 So, we can analyze the integer valued signal 
$k(t)$, instead of $\epsilon(t)$ \cite{Roux,Masl} (the 
 value of the smallest variable at time $t$), 
 in order to estimate numerically the exponent 
$\tau$. This alternative method
 allows us to avoid the problems 
 of the numerical approximations 
 that one faces when one analyzes
 the real signal $\epsilon(t)$. We
 get $\tau \simeq 1.60 \pm 0.02$ 
 (fig.\ref{fig3}), in very good agreement with 
our analytical calculation. From the knowledge of 
 the exponents $D_f$ and $\tau$  one can recover, 
 via scaling relations \cite{Masl}, 
 all the other critical exponents of IP.

 This new theoretical framework, discussed here
  in relation to the IP problem, 
 can easily be extended
 to other problems of the same extremal 
 nature like, for example, the
 Bak-Sneppen model \cite{BSRG}.

\begin{figure}
\protect\caption{Time evolution of the solution of the equation
for $\Phi_t(x)$ (for $2-d$ bond IP). $\Phi_t(x)$ 
tends asymptotically to a theta function with 
discontinuity at $p_c=1/2$.}
\label{figphi}
\end{figure}

\begin{figure}
\protect\caption{ Renormalization scheme for the extremal 
dynamics: (a)
 Dynamics at the smaller scale; (b) rescaled dynamics.}
\label{fig1}
\end{figure}

 \begin{figure}
\protect\caption{The distribution of critical avalanches
in invasion percolation. The solid line has slope $-1.60\pm0.02$.}
\label{fig3}
\end{figure}

\begin{table}
\begin{centering}
\protect\caption{FST results for IP without trapping ($D_{f}$),
with site trapping ($D_{f}^{I}$), with bond trapping 
($D_{f}^{II}$), and directed IP ($D_{f}^{DIP}$). The FST 
results are compared with known analytical and 
simulation values.}
\label{table}
\begin{tabular}{|c|c|c|c|c|}\hline
\rm{ORDER n } & $D_{f}(n)$ & $D_{f}^{I}(n)$ 
& $D_{f}^{II}(n)$ & $D_{f}^{DIP}(n)$ \\\hline\hline
$3$ & $1.7039$ & $1.6965$ & $1.7029$ & $1.6254$\\
$4$ & $1.7941$ & $1.7378$ & $1.7825$ & $1.6626$\\
$5$ & $1.8228$ & $1.7506$ & $1.8066$ & $1.6924$\\
$6$ & $1.8473$ & $1.7599$ & $1.8245$ & $1.7081$\\
$7$ & $1.8565$ & $1.7642$ & $1.8317$ & $1.7189$\\
$8$ & $1.8645$ & $1.7678$ & $1.8372$ & $1.7250$\\
$9$ & $1.8677$ & $1.7697$ & . . .   & . . .  \\
. . . & . . .  & . . .  & . . .  & . . .\\
$\infty$ & $1.8879$ & $1.7812$ & $1.8544$ & $1.7444$\\
\hline
$analyt.$ & $\frac{91}{48}\simeq1.895\tablenotemark[1]$ & $-$ & 
$-$ & $\simeq 1.748\tablenotemark[2]$\\
\hline
$simul.$ & $\sim1.89 \cite{Wilk}$ & $\sim1.82 \cite{Roux}$ 
& $\sim1.86 \cite{Roux}$ & $-$\\
\end{tabular}
\protect\tablenotetext[1] {conformal mapping 
applied to $2d$ Percolation \cite{Stauff}.}
\protect\tablenotetext[2] {series expansions \cite{Baxt}.}
\end{centering}
\end{table} 

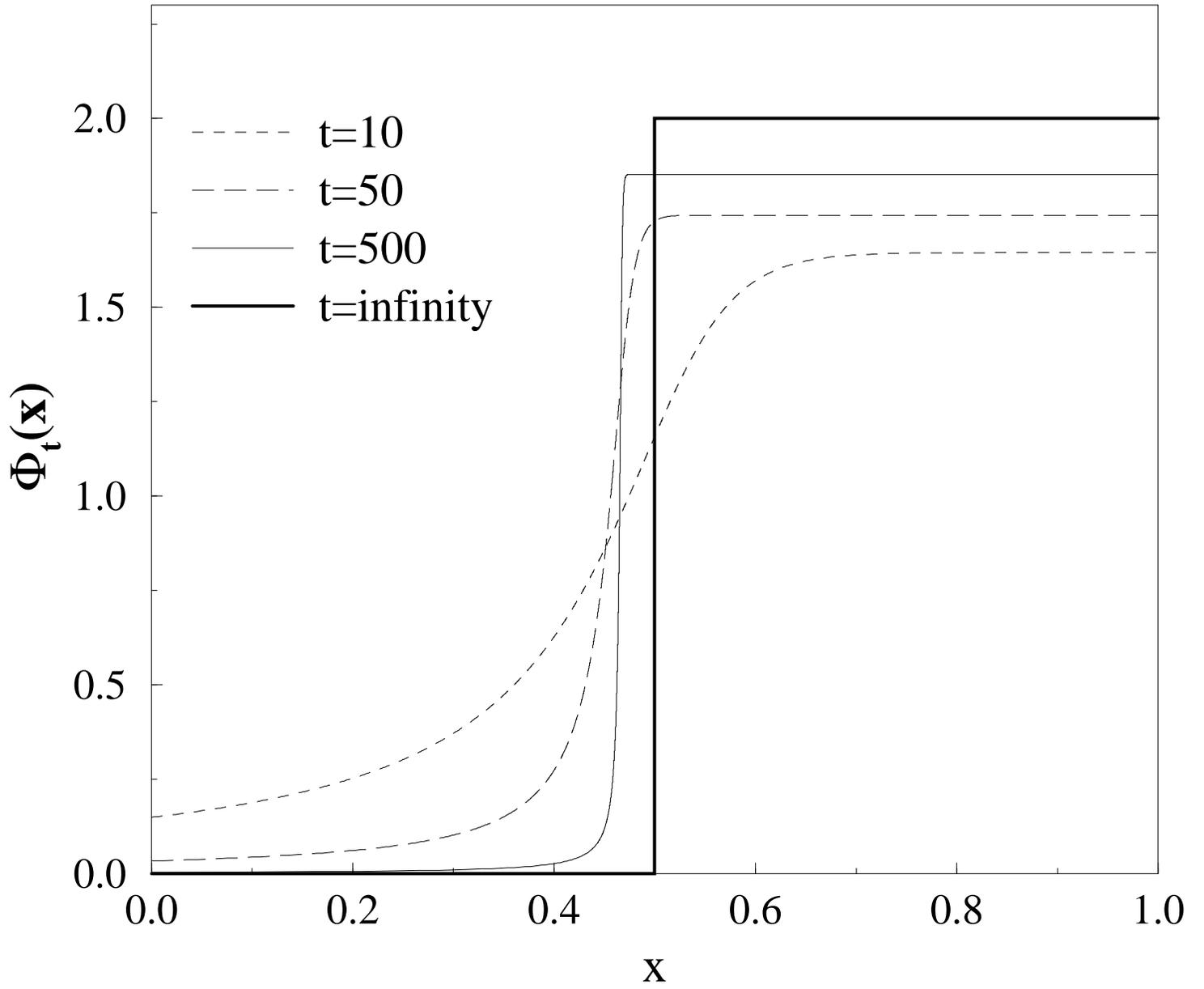
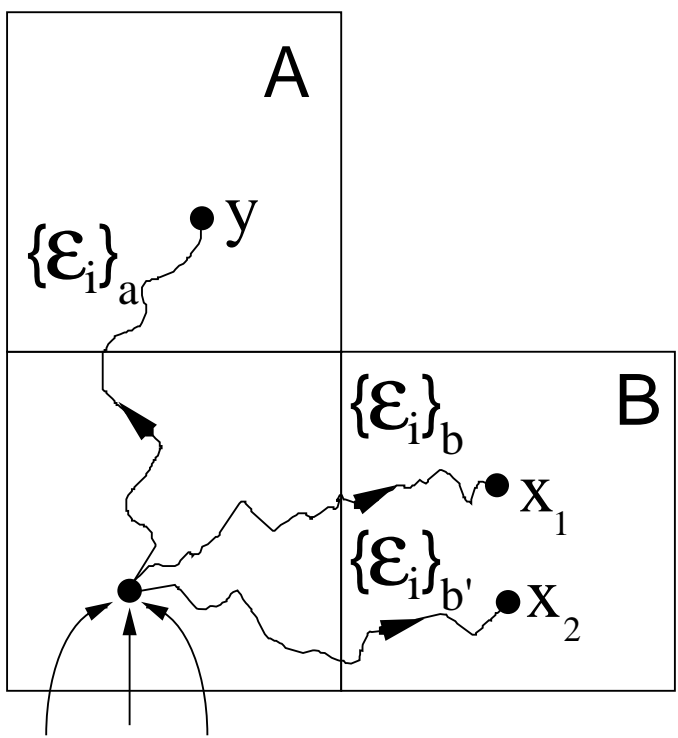
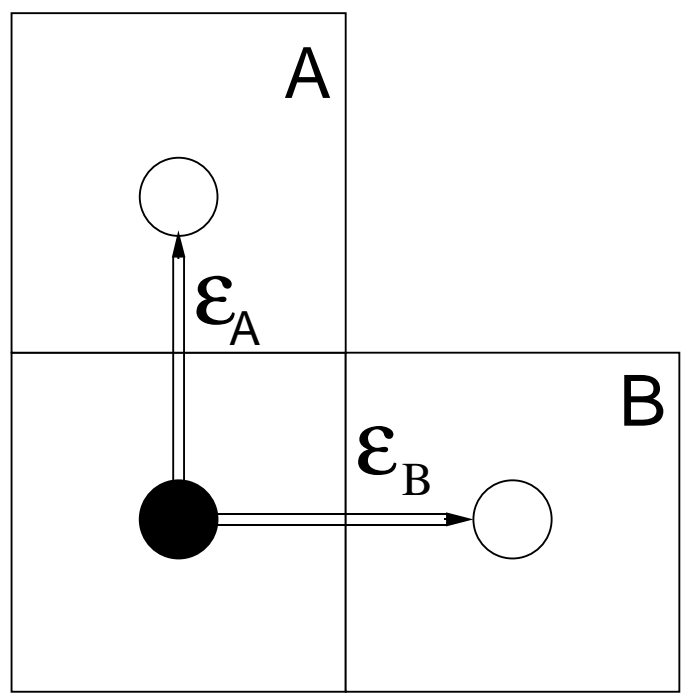
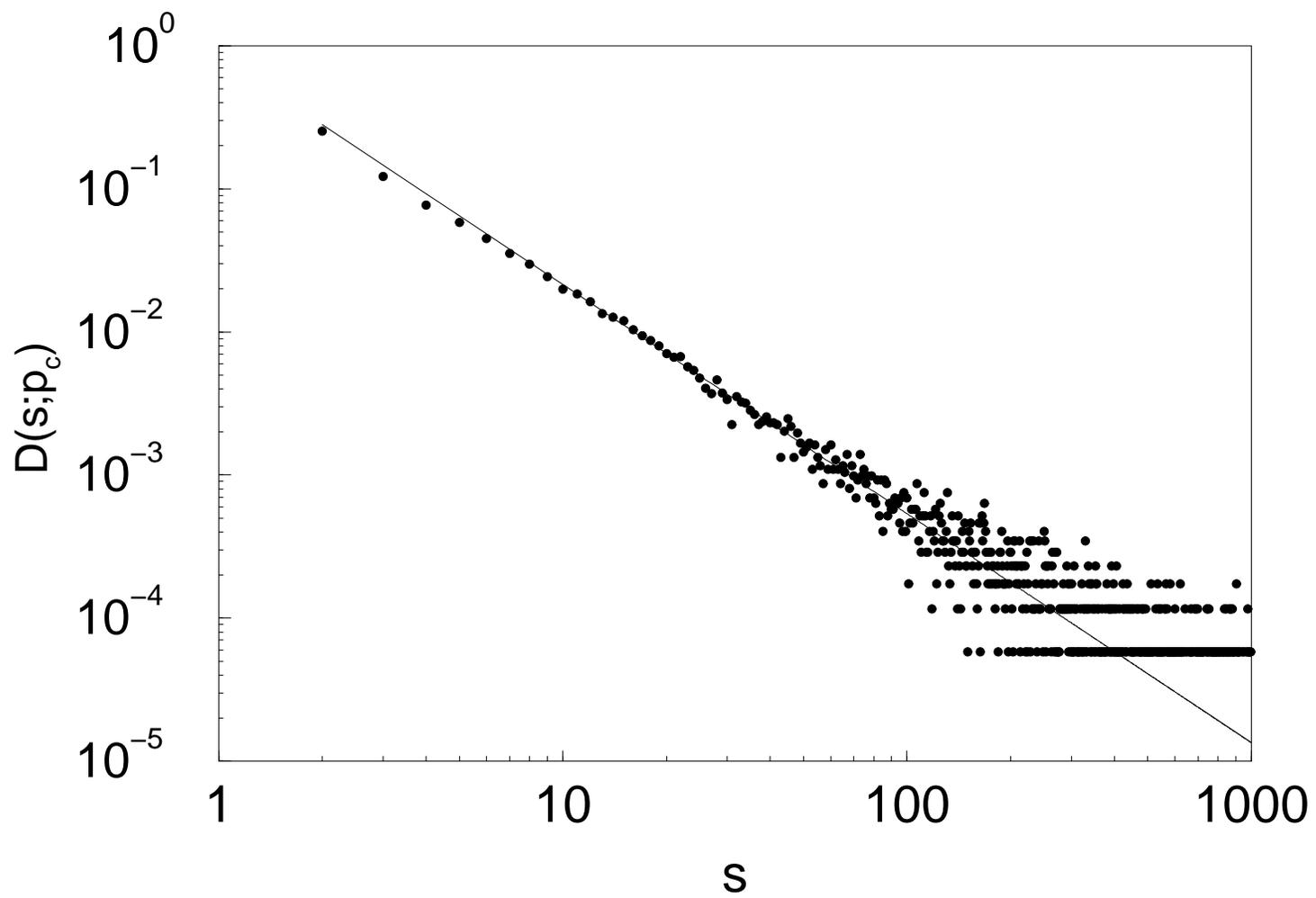
\end{document}